\documentclass[12pt,aps,prb,preprint]{revtex4}   

\usepackage{amsmath}    
\usepackage{amsfonts}  
\usepackage{amssymb}

\begin{document}

\title{PARITY AND THE SPIN-STATISTICS CONNECTION}

\author{J. A. MORGAN}
\address{The Aerospace Corporation, P. O. Box 92957 \\Los Angeles, CA 90009, 
United States of America \\
\email{john.a.morgan@aero.org}
}


\begin{abstract}
The spin-statistics connection is obtained in a simple and elementary way for general causal 
fields by using the parity operation to exchange spatial coordinates in the scalar product of a 
locally commuting field operator, evaluated at position 
$\mbox{\bf{x}}$, with the same field operator evaluated at $-\mbox{\bf{x}}$, 
at equal times.

\end{abstract}


\maketitle

\section{Introduction}
Proofs of the spin-statistics theorem tend, broadly speaking, to fall into two classes.  The
first class, historically, depends upon analytic properties of field operator 
commutators.\cite{Pauli1940,B1958,LZ1958,StreaterWightman1964}  The second class invokes 
topological arguments.  Proofs in this latter class variously use homotopies in configuration 
space for identical particles\cite{RF1968,Tsch1989,Tsch1990,Balach1990,Balach1993} or arguments 
involving adiabatic exchange of particles carrying topological 
markers.\cite{Feynman1987,BR1997}  
The proof by Schwinger\cite{Schwinger1951} stands apart from both classes 
in exploiting the discrete symmetry of time-reversal.  

The use, on the one hand, of the 
exchange of identical particles in the topological 
theorems, and, on the other, of a discrete symmetry applied to a scalar
invariant (the Lagrangian density of a field) in Schwinger's proof, suggests using another discrete 
symmetry, parity, to examine the effect of exchanging particle coordinates by passive 
transformations.  This note presents 
a simple demonstration of the spin-statistics connection based upon that idea.  
The proof is elementary, and in essence algebraic.

\section{Parity and Causal Fields}
Irreducible representations of the Poincar\'{e} group are classified according to eigenvalues
of two angular momentum-like infinitesimal generators $\mbox{\bf{A}}$ and 
$\mbox{\bf{B}}$.\cite{Pauli1940,StreaterWightman1964,Weinberg1995,Weinberg1969,Tung1985}  The 
(A,B) representation contains multiple spin angular momentum quantum numbers $|A-B| \le j \le A+B$.  
General 
fields are built up from the $(A,B)$ representations.  Familiar examples
include the $(0,0)$ scalar field, and the $(\frac{1}{2},0) \oplus (0,\frac{1}{2})$ Dirac field.

Let a spin j massive field $\Psi^{(AB)}$ be an element of a given (A,B) 
representation.  The construction of this object is given in Refs.~\cite{Weinberg1969},
\cite{Weinberg1995b}.  Applying the parity operation $P$  
gives\cite{Weinberg1995}
\begin{equation}
P \Psi^{(AB)}_{ab}(\mbox{\bf{x}},t) P^{-1} =
\eta_{P} (-1)^{A+B-j} \Psi^{(BA)}_{ba}(-\mbox{\bf{x}},t) \label{eq:eqn1}
\end{equation}
The intrinsic parity $\eta_{P}$ of the field is $\pm{1}$.  
The action of $P$ has no effect on spin, and assumes nothing regarding 
statistics. 

\section{Spin and Statistics:  Weinberg Fields} 
Before considering the general case, the method of proof is worked out for the
simpler case of $(j,0)$ representations, sometimes called Weinberg fields.\cite{Weinberg1964}  
Define the field operator
\begin{equation}
\xi_{\sigma} \equiv \Psi^{(j0)}_{\sigma}
\end{equation}
where $\sigma$ runs from $-j$ to $j$.
The field $\xi_{\sigma}(x)$ annihilates a spin j particle (or creates an antiparticle) localized at 
spacetime point $x$, with z-projection of angular momentum $\sigma$.

It will be shown that imposing local commutativity on $\xi$
leads to the spin-statistics connection.  Consider the field $\xi$ evaluated at two points in 
spacetime 
separated by spacelike interval.  A Lorentz frame exists in which the two 
points occur at equal time, 
so we may write the fields as $\xi(\mbox{\bf{x}},t)$ and $\xi(-\mbox{\bf{x}},t)$.  The effect of
$P$ on their scalar product is, according to Eq (\ref{eq:eqn1}) for $A=B=0$, 
\begin{eqnarray}
 P \xi (\mbox{\bf{x}},t) \cdot \xi (\mbox{-\bf{x}},t) P^{-1} \nonumber \\
=P \xi (\mbox{\bf{x}},t) P^{-1} \cdot P \xi (\mbox{-\bf{x}},t) P^{-1} \nonumber \\
=\xi (-\mbox{\bf{x}},t) \cdot \xi (\mbox{\bf{x}},t)\label{eq:sproduct1}
\end{eqnarray}
Equation (\ref{eq:sproduct1}) is the product of 
two quantities with the same parity, and is thus an even parity scalar function of $\bf{x}$.  
Considered as a function of $\mbox{\bf{x}}$, an even parity scalar operator obeys
$P f(\mbox{\bf{x}}) P^{-1}=f(\mbox{\bf{x}})$, thus
\begin{equation}
\xi (\mbox{\bf{x}},t) \cdot \xi (-\mbox{\bf{x}},t)
=\xi (-\mbox{\bf{x}},t) \cdot \xi (\mbox{\bf{x}},t).  \label{eq:sproduct2p5}
\end{equation}

The product on the right-hand side of Eq (\ref{eq:sproduct2p5}) is the scalar product of 
two irreducible spherical tensors of the same rank.  It is given 
by\cite{Edmonds1960,Racah1942}
\begin{equation}
\xi(-\mbox{\bf{x}},t) \cdot \xi (\mbox{\bf{x}},t)=
\sum_{\sigma}(-1)^{\sigma}\xi_{\sigma}(-\mbox{\bf{x}},t) \label{eq:scalarprod}
\xi_{-\sigma}(\mbox{\bf{x}},t).  
\end{equation}
By hypothesis, commutation relations of a causal field ($-$ for Bose, $+$ for Fermi) 
vanish outside the light cone; in particular\cite{footnote1}
\begin{equation}
[\xi_{\sigma}(\mbox{\bf{x}},t),\xi_{\lambda}(-\mbox{\bf{x}},t)]_{\mp}=0 \label{eq:ccr}
\end{equation}
Therefore,
\begin{equation}
\xi(-\mbox{\bf{x}},t) \cdot \xi (\mbox{\bf{x}},t)=
\pm\sum_{\sigma}(-1)^{\sigma}\xi_{-\sigma}
(\mbox{\bf{x}},t)\xi_{\sigma}(-\mbox{\bf{x}},t), \label{eq:sproduct4}
\end{equation}
as the fields are Bose or Fermi.  Upon inverting the order of summation by replacing
$\sigma$ with $-\sigma'$, 
\begin{equation}
\xi(-\mbox{\bf{x}},t) \cdot \xi (\mbox{\bf{x}},t)=
\pm\sum_{\sigma'}(-1)^{-\sigma'}\xi_{\sigma'}
(\mbox{\bf{x}},t)\xi_{-\sigma'}(-\mbox{\bf{x}},t),
\end{equation}
and noting
\begin{equation}
(-1)^{-\sigma'}= \left\{ \begin{array}{cc}
(-1)^{\sigma'} & \mbox{integer $j$} \\
-(-1)^{\sigma'} & \mbox{half-integer $j$}
\end{array}
\right.
=(-1)^{2j}(-1)^{\sigma'},
\end{equation}
we obtain for Eq (\ref{eq:sproduct2p5})
\begin{equation}
\xi(\mbox{\bf{x}},t) \cdot \xi (-\mbox{\bf{x}},t)=
\pm(-1)^{2j}\xi (\mbox{\bf{x}},t) \cdot \xi (-\mbox{\bf{x}},t). \label{eq:sproduct5}
\end{equation}

Take the matrix element of both sides of Eq (\ref{eq:sproduct5}) between the vacuum and 
a state with one 
quantum of the field
$\xi$ localized at $\mbox{\bf{x}}$ with z-value of its spin equal to $\rho$ and one quantum 
at $\mbox{-\bf{x}}$, with spin z-value $-\rho$.  Eq (\ref{eq:sproduct5}) becomes
\begin{eqnarray}
\lefteqn{\langle  VAC| \xi_{\rho}(\mbox{\bf{x}},t) 
\xi_{-\rho}(-\mbox{\bf{x}},t) | 
(+\mbox{\bf{x}},t; +\rho) (-\mbox{\bf{x}},t; -\rho) \rangle =} \nonumber \\
& & \pm(-1)^{2j} \langle VAC 
| \xi_{\rho}(\mbox{\bf{x}},t) \xi_{-\rho} (-\mbox{\bf{x}},t) |
 (+\mbox{\bf{x}},t; +\rho) (-\mbox{\bf{x}},t;-\rho) \rangle \label{eq:me}
\end{eqnarray}
By hypothesis, a $\rho$ exists for which the matrix element is nonvanishing, allowing us to 
conclude 
\begin{equation}
1=\pm(-1)^{2j}, \label{eq:drumroll}
\end{equation} 
which is the connection between spin and statistics.

\section{Spin and Statistics:  General Fields}
The argument just given is readily extended to the case of the general $(A,B)$ representation.  
The field $\xi^{(AB)}_{mn}$ now carries two indices $-A \le m \le A$ and
$-B \le n \le B$, and the scalar product in Eq (\ref{eq:scalarprod}) is replaced by an 
expresson that couples two $(A,B)$ spherical tensors to a $(0,0)$ scalar, in an extension 
of Racah's\cite{Racah1942} original derivation of Eq (\ref{eq:scalarprod}), which now 
becomes (retaining the dot product notation)
\begin{eqnarray}
\sum_{m,n} \left( \begin{array}{rlc}
A & A & 0 \\
-m & m & 0 
\end{array} \right)
\left( \begin{array}{rlc}
B & B & 0 \\
-n & n & 0 
\end{array} \right) 
\xi_{mn}(-\mbox{\bf{x}},t) \xi_{-m-n}(\mbox{\bf{x}},t) \nonumber \\
\propto \sum_{m,n}(-1)^{\sigma}\xi_{mn}(-\mbox{\bf{x}},t) 
\xi_{-m-n}(\mbox{\bf{x}},t) \nonumber \\
\equiv \xi(-\mbox{\bf{x}},t) \cdot \xi (\mbox{\bf{x}},t)  \label{eq:newsproduct}
\end{eqnarray}
where $\sigma=m+n$, and the objects in parentheses are Wigner 3j symbols.  
By Eq (\ref{eq:eqn1}) for the (0,0) representation, the result of applying $P$ to 
Eq (\ref{eq:newsproduct}) once again gives Eq (\ref{eq:sproduct2p5}). 
Both the spin $j$ and summation index $\sigma$ are half-integral if and only if one of 
$A$ and $B$ is half-integral.  Therefore, Eq (\ref{eq:sproduct5}) holds for the general $(A,B)$
representation, and taking the matrix element of 
Eq (\ref{eq:sproduct5}) between the vacuum and 
a suitable state 
$| (\mbox{\bf{x}},t; \mu,\nu) (-\mbox{\bf{x}},t; -\mu,-\nu) \rangle$ gives, again,
the proper spin-statistics connection.


\begin{thebibliography}{5}

\bibitem{Pauli1940} W. Pauli, ``The connection between spin and statistics,'' 
Phys. Rev. {\bf 58}, 716--722 (1940).

\bibitem{B1958}N. Burgoyne, ``On the connection of spin and statistics,'' Nuovo
Cimento {\bf8}, 607--609 (1958).

\bibitem{LZ1958}Gerhard L\"{u}ders, and Bruno Zumino, ``Connection between spin
and statistics,'' Phys. Rev. {\bf110}, 1450--1453 (1958).

\bibitem{StreaterWightman1964}R. F. Streater and A. S. Wightman, \textit{PCT, Spin and Statistics, 
and All That} (W. A. Benjamin, New York, NY, 1964).

\bibitem{RF1968}Julio Finkelstein and David Rubinstein, ``Connection between
spin, statistics, and kinks,'' J. Math. Phys. {\bf 9}, 1762--1779 (1968). 

\bibitem{Tsch1989}Ralf D. Tscheuschner, ``Topological spin-statistics relation
in quantum field theory,'' Int. J. Theor. Phys. {\bf28}, 1269--1310 (1989).

\bibitem{Tsch1990}Ralf D. Tscheuschner, ``Coinciding versus noncoinciding: Is
the topological spin-statistics theorem already proven in quantum mechanics?,''
J. Math. Phys. {\bf32}, 749--752 (1990).

\bibitem{Balach1990}A. P. Balachandran, A. Daughton, Z.-C. Gu, G. Marmo, R. D.
Sorkin, and A. M. Srivastava, ``A topological spin-statistics theorem or a use
of the antiparticle,'' Mod. Phys. Lett. A{\bf 5}, 1575--1585 (1990).

\bibitem{Balach1993}A. P. Balachandran, A. Daughton, Z.-C. Gu, R. D. Sorkin, G.
Marmo, and A. M. Srivastava, ``Spin-statistics theorems without relativity or
field theory,'' Int. J. Mod. Phys. A{\bf 8}, 2993--3044 (1993).

\bibitem{Feynman1987}R. P. Feynman, ``The reason for antiparticles,'' in 
R. P. Feynman and S. Weinberg, \textit{Elementary Particles and the Laws of Physics} 
(Cambridge University Press, 1987). 

\bibitem{BR1997}M. V. Berry and J. M. Robbins, ``Indistinguishability for
quantum particles: Spin, statistics, and the geometric phase," Proc. Roy. Soc.
Lond. A{\bf 453}, 1771--1790 (1997).

\bibitem{Schwinger1951} Julian Schwinger, ``Theory of quantized fields I,'' Phys
Rev. {\bf 82}, 914--927 (1951).



\bibitem{Weinberg1995}Steven Weinberg, \textit{The Quantum Theory of Fields I.} (Cambridge 
University Press, Cantab., 1995) pp. 239-240

\bibitem{Weinberg1969}Steven Weinberg, "Feynman Rules for Any Spin III", Phys. Rev. {\bf181}, 
1893-1899 (1969)

\bibitem{Tung1985}W.-K.Tung, \textit{Group Theory in Physics} (World Scientific,
Singapore, 1985), Chaps. 7--10.

\bibitem{Weinberg1995b}Ref.~\onlinecite{Weinberg1995}, pp. 233-243

\bibitem{Weinberg1964}Steven Weinberg, "Feynman Rules for Any Spin", Phys. Rev. {\bf133B}, 
1318-1332 (1964)

\bibitem{Edmonds1960}A. R. Edmonds, \textit{Angular Momentum in Quantum
Mechanics} (Princeton University Press, Princeton, 1960), p. 72.

\bibitem{Racah1942}G. Racah, "Theory of complex spectra II", Phys. Rev. {\bf62}, 438-462 (1942)

\bibitem{footnote1} It suffices to consider commutation relations between the fields, rather
that the usual relations between a field and its Hermitian conjugate, \emph{vide.} 
G. F. Dell'Antonio, ``On the connection between spin and
statistics,'' Ann. Phys. {\bf16}, 153--157 (1961).  

\bibitem{Weinberg1995c}Ref.~\onlinecite{Weinberg1995}, pp. 229-233

\end{thebibliography}
\end{document}